\documentclass[aip, apl,reprint,twocolumn,superscriptaddress]{revtex4-1}%
\usepackage{graphicx,color,dcolumn,bm}
\usepackage[colorlinks,linkcolor=blue,anchorcolor=blue,citecolor=blue,urlcolor=blue]%
{hyperref}
\usepackage{amsmath}
\usepackage{amsfonts}
\usepackage{amssymb}

\begin{document}
	\title{Semiconductor-to-metal transition in bilayer MoSi$_{2}$N$_{4}$ and WSi$_{2}$N$_{4}$ with strain and
		electric field}
	\author{Qingyun Wu}
	\email{qingyun\_wu@sutd.edu.sg}
	\affiliation{Science, Mathematics and Technology,
		Singapore University of Technology and Design (SUTD), 8 Somapah Road,
		Singapore 487372, Singapore}
	\author{Liemao Cao}
	\affiliation{College of Physics and Electronic Engineering, Hengyang Normal University, Hengyang 421002, China}
	\author{Yee Sin Ang}
	\affiliation{Science, Mathematics and Technology,
		Singapore University of Technology and Design (SUTD), 8 Somapah Road,
		Singapore 487372, Singapore}
	\author{Lay Kee Ang}
	\email{ricky\_ang@sutd.edu.sg}
	\affiliation{Science, Mathematics and Technology,
		Singapore University of Technology and Design (SUTD), 8 Somapah Road,
		Singapore 487372, Singapore}
	
	\begin{abstract}
		With exceptional electrical and mechanical properties and at the same time
		air-stability, layered MoSi$_{2}$N$_{4}$ has recently draw great attention.
		However, band structure engineering via strain and electric field, which is
		vital for practical applications, has not yet been explored. In this work, we show
		that the biaxial strain and external electric field are effective ways for the
		band gap engineering of bilayer MoSi$_{2}$N$_{4}$ and WSi$_{2}$N$_{4}$. It is
		found that strain can lead to indirect band gap to direct band gap
		transition. On the other hand, electric field can result in semiconductor to
		metal transition. Our study provides insights into the band structure
		engineering of bilayer MoSi$_{2}$N$_{4}$ and WSi$_{2}$N$_{4}$ and would pave the
		way for its future nanoelectronics and optoelectronics applications.
		
	\end{abstract}
	\maketitle

	Since the discovery of
	graphene,\cite{novoselov2004electric,novoselov2005two,novoselov2005two1} there
	has been great research interests in two-dimensional (2D) layered materials
	due to their exceptional fundamental physical properties and tremendous
	potentials in devices applications.\cite{bhimanapati2015recent,tan2017recent}
	Owing to their high carrier mobility, excellent on/off ratio and unique
	optoelectronic properties, 2D transition metal dichalcogenides
	(TMD)\cite{mak2010atomically,radisavljevic2011single,splendiani2010emerging}
	and black
	phosphorene\cite{liu2014phosphorene,li2014black,koenig2014electric,qiao2014high}
	have been widely studied in last decade. To fully utilized those
	semiconducting 2D layered materials in nanoelectronics and optoelectronic
	devices, band gap engineering is indispensable and plays an essential role in
	device applications. One of the effective ways of tuning band gap is to apply
	strain to the layered materials. By doing so, electronic properties of TMD and
	phosphorene can be altered and strain sensitive device can be
	designed and fabricated.\cite{johari2012tuning,yun2012thickness,bhattacharyya2012semiconductor,pan2012tuning,guo2014phosphorene,rodin2014strain,peng2014strain,fei2014strain}
	The other strategy to achieve band gap engineering is by applying an external
	electric field, which may reduce the band gap for bilayer system and
	even lead to semiconductor to metal
	transition.\cite{ramasubramaniam2011tunable,liu2012tuning,chu2015electrically,dolui2012electric,kou2012tuning,liu2015switching,wu2015electronic,huang2015electric,cao2019janus}

	\begin{figure}[ptb]
		\centering
		\includegraphics[width=0.5\textwidth]{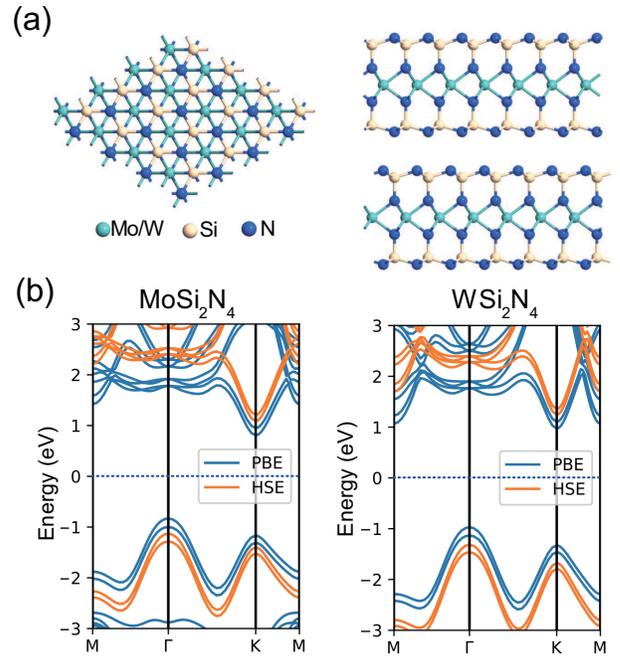}
		\caption{(a) Top and side view of bilayer MoSi$_{2}$N$_{4}$ and WSi$_{2}$N$_{4}$.
			(b) PBE and HSE band structure of bilayer MoSi$_{2}$N$_{4}$ and WSi$_{2}$N$_{4}
			$.}%
		\label{fig1}%
	\end{figure}

	Most recently, layered 2D MoSi$_{2}$N$_{4}$ and WSi$_{2}$N$_{4}$ has been
	experimentally synthesized using chemical vapor deposition
	(CVD).\cite{hong2020chemical} Experimental results suggest that they have
	semiconducting characteristic with good mechanical strength and air-stability.
	The paper also reported an intrinsic electron and hole mobilities of 270
	cm$^{2}$V$^{-1}$s$^{-1}$ and 1200 cm$^{2}$V$^{-1}$s$^{-1}$, respectively, for layered 2D MoSi$_{2}$N$_{4}$,  
	which is around four times as that of MoS$_{2}$.\cite{hong2020chemical} With
	higher carrier mobility than MoS$_{2}$ and better air-stability than
	phosphorene, MoSi$_{2}$N$_{4}$ and its family structure of MA$_{2}$Z$_{4}$
	monolayers (M represents an early transition metal, A is Si or Ge, Z stands
	for N, P or As) have soon received much
	attention.\cite{novoselov2020discovery,tang2020chemical,li2020valley,cao2021two,guo2020coexistence,guo2020intrinsic,mortazavi2020exceptional,bafekry2020mosi2n4,yang2020valley,yu2020high,kang2020second,wang2020unexpected}
	Nevertheless, there is still lack of band gap engineering study of bilayer
	MoSi$_{2}$N$_{4}$ and WSi$_{2}$N$_{4}$, which is critical for the future
	application of \ layered MA$_{2}$Z$_{4}$ material in nanoelectronics and
	optoelectronic devices up to date.
	
	In this work, we explore the tuning of electrical band structure of bilayer
	MoSi$_{2}$N$_{4}$ and WSi$_{2}$N$_{4}$ under strain and external electric field.
	We find that band gaps can be tuned by both compress and tensile strain, which is a result of the change in Mo/W-N bond length and
	Mo-Mo/W-W distance induced by strain. We also find that the electric field can
	modulate the band gap, due to the charge
	redistribution induced by the electric field. Our findings imply that the
	strain and electric field tunable bilayer MoSi$_{2}$N$_{4}$ and WSi$_{2}$N$_{4}$
	could be promising materials for the next generation nanoelectronics and optoelectronics.\

	\begin{figure*}[ptb]
		\centering
		\includegraphics[width=0.9\textwidth]{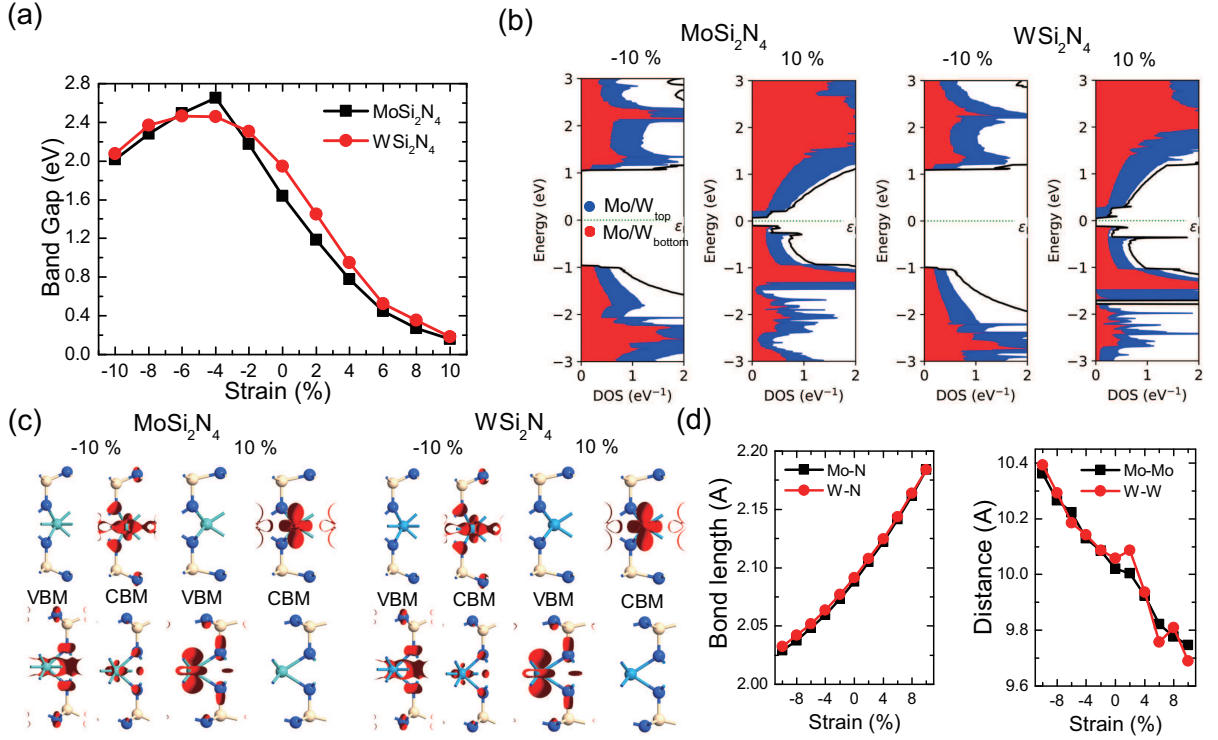}
		\caption{(a) Band gap as a function of applied strain for bilayer MoSi$_{2}%
			$N$_{4}$/WSi$_{2}$N$_{4}$. (b) Projected density of states (PDOS) for bilayer
			MoSi$_{2}$N$_{4}$ and WSi$_{2}$N$_{4}$ under -10 \% and 10 \% strain. (c) Partial
			charge density corresponding to the valance band maximum (VBM) and conduction
			band minimum (CBM) for bilayer MoSi$_{2}$N$_{4}$ and WSi$_{2}$N$_{4}$ under -10 \%
			and 10 \% strain. (d) Mo-N/W-N bond length and Mo-Mo/W-W distance as a
			function of applied strain for bilayer MoSi$_{2}$N$_{4}$ and WSi$_{2}$N$_{4}$.}%
		\label{fig2}%
	\end{figure*}

	Our first-principles study is carried out based on the density functional
	theory (DFT) calculations. The Vienna \textit{Ab initio} Simulation Package
	(VASP)\cite{vasp1,vasp2} was adopted for the geometry optimization and the
	QuantumATK\cite{smidstrup2019quantumatk} was used to investigate the
	electronic properties of bilayer MoSi$_{2}$N$_{4}$ and WSi$_{2}$N$_{4}$ under
	strain and external electric field. A gamma-centered Brillouin zone $k$-point
	sampling grid using the Monkhorst-Pack method\cite{mp} of $9\times9\times1$
	was used for geometry optimization and a grid of $15\times15\times1$ for
	property calculations. Atomic geometry optimization criteria is that all
	forces are smaller than 0.01 eV/\AA . The generalized gradient approximation
	(GGA) with the Perdew-Burke-Ernzerhof form (PBE)\cite{pbe} was chosen for the
	exchange-correlation functional in the calculations. The
	Heyd-Scuseria-Ernzerhof hybrid functional method (HSE)\cite{heyd2003hybrid}
	was also used to obtain the more accurate band structure of the bilayers. To
	take into account the weak van der Waals interactions in the bilayers, we
	adopted the DFT-D3 method with the Grimme scheme\cite{Grimme2010, Grimme2011}
	in the calculations. A 20 \AA \ thick of vacuum layer was inserted between
	adjacent bilayers to eliminate the interactions from periodic images.\cite{wu2019design}
	
	Based on our total energy calculations and also previous
	results,\cite{hong2020chemical} the AC stacking of the bilayer MoSi$_{2}%
	$N$_{4}$ and WSi$_{2}$N$_{4}$ is the most stable bilayer configuration with the
	lowest total energy. Therefore, we only focus on the AC stacking in the
	following discussions. Figure~\ref{fig1}(a) depicts the optimized geometric
	structure of the bilayer MoSi$_{2}$N$_{4}$ and WSi$_{2}$N$_{4}$. Our calculated
	lattice parameter of the bilayer MoSi$_{2}$N$_{4}$ and WSi$_{2}$N$_{4}$ is
	$a=2.90/2.90$ \AA , which is in good agreement with previous calculated
	results.\cite{hong2020chemical} As shown in Fig.~\ref{fig1}(b) the band structure
	of the bilayer MoSi$_{2}$N$_{4}$ and WSi$_{2}$N$_{4}$ is very similar to its
	monolayer counterpart with only small splitting in energy bands induced by the
	interlayer interactions, which slightly reduces the band gap. The indirect
	band gap of the bilayer MoSi$_{2}$N$_{4}$ and WSi$_{2}$N$_{4}$ is calculated to be
	1.64/1.94 eV. To obtain a more accurate band structure, the HSE hybrid
	functional calculations were also performed. It is found that the shape of the
	HSE bands are similar to that of the PBE results with only valence bands
	pushed down and conduction bands pushed up. We obtain the band gap value of
	2.22/2.56 eV for the bilayer MoSi$_{2}$N$_{4}$ and WSi$_{2}$N$_{4}$ in HSE calculations.
	
	\begin{figure*}[ptb]
		\centering
		\includegraphics[width=1.0\textwidth]{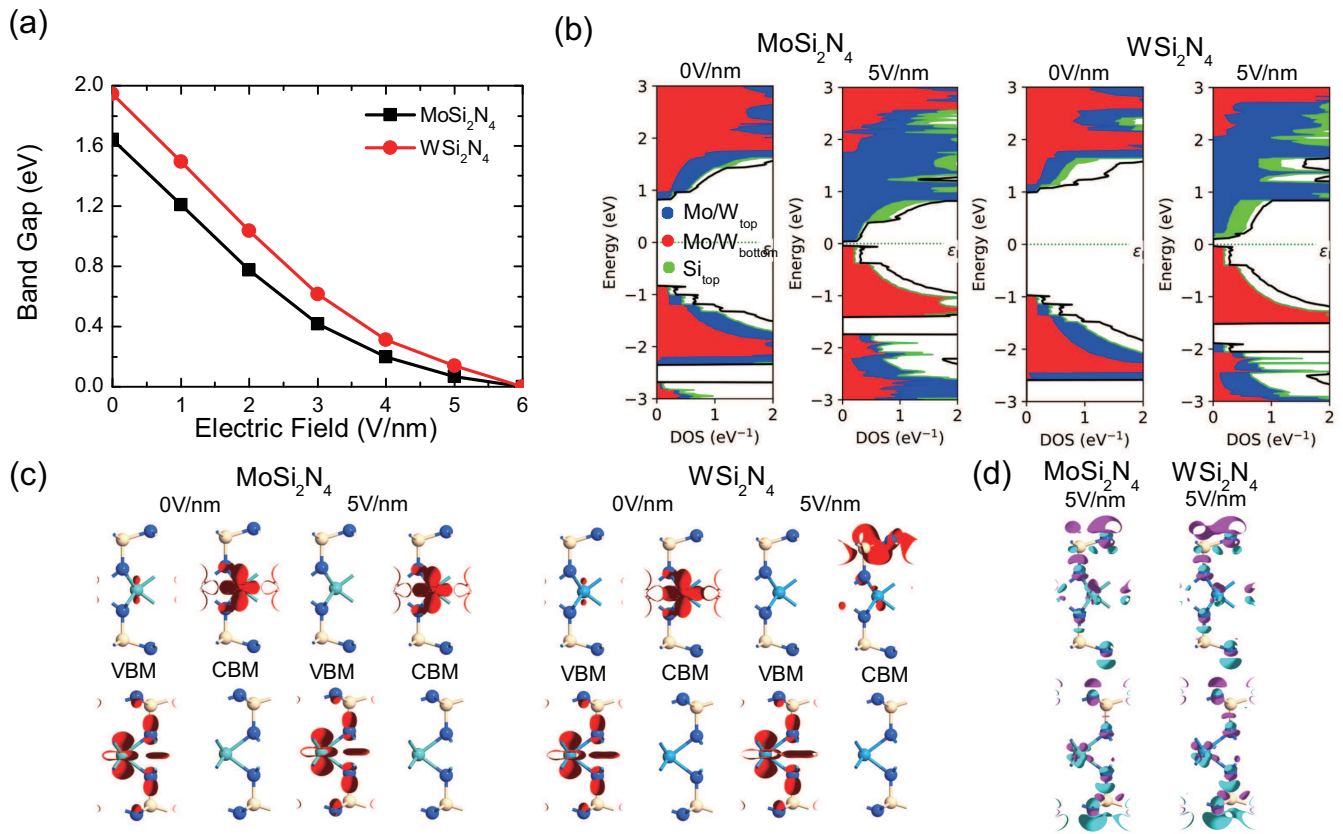}
		\caption{(a) Band gap as a function of applied electric field for bilayer
			MoSi$_{2}$N$_{4}$ and WSi$_{2}$N$_{4}$. (b) Projected density of states (PDOS) for
			bilayer MoSi$_{2}$N$_{4}$ and WSi$_{2}$N$_{4}$ under 0 V/nm and 5 V/nm electric
			field. (c) Partial charge density corresponding to the valance band maximum
			(VBM) and conduction band minimum (CBM) for bilayer MoSi$_{2}$N$_{4}$%
			/WSi$_{2}$N$_{4}$ under 0 V/nm and 5 V/nm electric field. (d) Charge
			redistribution for bilayer MoSi$_{2}$N$_{4}$ and WSi$_{2}$N$_{4}$ under 5 V/nm
			electric field. The blue (red) isosurfaces represent charge accumulation
			(depletion).}%
		\label{fig3}%
	\end{figure*}

	It has been proved that strain engineering is an effective avenue to modulate
	the electric, magnetic and optical properties of 2D layered materials. To
	understand the tunability of the energy band gap of the bilayer MoSi$_{2}%
	$N$_{4}$ and WSi$_{2}$N$_{4}$, we applied biaxial strain to the system. The strain
	($\varepsilon$) is evaluated using $\varepsilon=(a-a_{0})/a_{0}\times100\%$,
	where $a$ and $a_{0}$ are the lattice parameter of the strained and unstrained
	bilayer, respectively. Figure~\ref{fig2}(a) shows the evolution of the band
	gap as a function of applied strain for bilayer MoSi$_{2}$N$_{4}$ and WSi$_{2}%
	$N$_{4}$. It is suggested that the band gap undergoes a monotonous decreasing
	with the increasing tensile strain. However, with increasing compress strain,
	the band gap increases then decreases at around -4 \% for the bilayer
	MoSi$_{2}$N$_{4}$\ and -6 \% for the bilayer MoSi$_{2}$N$_{4}$. Besides
	varying the band gap, more importantly, strain is also found to
	change an indirect band gap semiconductor to a direct band gap semiconductor
	at about -4 \% of compress strain for the bilayer MoSi$_{2}$N$_{4}%
	$.(See Fig.~S1) To understand the electronic structure engineering by strain, we
	draw the projected density of states (PDOS) and partial charge density
	corresponding to the valance band maximum (VBM) and conduction band minimum
	(CBM) for bilayer MoSi$_{2}$N$_{4}$ and WSi$_{2}$N$_{4}$ under -10 \% and 10 \%
	strain in Fig.~\ref{fig2}(b) and Fig.~\ref{fig2}(c). As can be seen from the figure,
	the VBM and CBM are mostly contributed from both the metal atoms of the top
	and bottom layer of the bilayer with an enlarged band gap for the compress -10
	\% strain in the PDOS. In the partial charge density plot, we see more
	d$_{xy}$ and d$_{x^{2}-y^{2}}$ orbitals\ for the metal atoms in VBM and CBM
	for the compress -10 \% strain. This is because the d$_{xy}$ and
	d$_{x^{2}-y^{2}}$ orbitals are favorable for the in-plane bonding. The reduced
	bond length of Mo/W-N under compress strain facilitate the in-plane bonding,
	as shown in Fig.~\ref{fig2}(d). The VBM is mostly from the Mo/W atom of the
	bottom layer and the CBM is mostly from the Mo/W atom of the top layer,
	respectively, with a reduced band gap for the tensile 10 \% strain. This split
	of the VBM and CBM from the metal atoms of the two different layer can be
	understood from the shorten distance of two metal atoms which increases the
	interlayer interactions as shown in Fig.~\ref{fig2}(d). Different from the
	compressed case, we see more d$_{z^{2}}$ orbitals\ for the metal atoms in VBM
	and CBM for the tensile 10 \% strain. This is because the d$_{z^{2}}$ orbital
	is favorable for the out-of-plane bonding. When the bond length increase, the
	out-of-plane bonding become more significant and thus more
	d$_{z^{2}}$ orbitals.
	
	Applying an external electric field perpendicular to the layered material is
	another effective way of tuning the properties of 2D layered materials. The evolution of the
	band gap as a function of the applied electric field for bilayer MoSi$_{2}%
	$N$_{4}$ and WSi$_{2}$N$_{4}$ is provided in Fig.~\ref{fig3}(a). It is found that
	the band gap decreases monotonously with the increasing electric field and
	closes at 6 V/nm for both bilayer MoSi$_{2}$N$_{4}$ and WSi$_{2}$N$_{4}$. It is
	also interesting to find that the CBM for MoSi$_{2}$N$_{4}$\ is at the K point
	while the CBM for WSi$_{2}$N$_{4}$ moves to the M point when the band gap
	closes.(See Fig.~S2) To gain more insight to the electronic structure engineering
	by electric field, we plot the PDOS and partial charge density of VBM and CBM
	for bilayer MoSi$_{2}$N$_{4}$ and WSi$_{2}$N$_{4}$ under 0 V/nm and 5 V/nm
	electric field in Fig.~\ref{fig3}(b) and Fig.~\ref{fig3}(c). As can be seen from the
	figure, the overall shape of the PDOS of the bottom and top metal atoms almost
	remain the same under 5V/nm electric field. However, the energy bands of the
	bottom metal atom shift up while the energy bands of the up atom shift down,
	which gradually closes the band gap. Furthermore, the CBM of the bilayer
	WSi$_{2}$N$_{4}$ under 5V/nm electric field comes from the top Si atom that
	can be observed both from the PDOS and the partial charge density plot. This
	explains why the CBM for WSi$_{2}$N$_{4}$ moves to the M point under 5V/nm
	electric field. This band gap engineering by external electric field can be
	interpreted from the charge redistribution of the bilayer MoSi$_{2}$N$_{4}%
	$ and WSi$_{2}$N$_{4}$ under electric field as shown in Fig.~\ref{fig3}(d). The
	charge density difference ($\Delta\rho$) is defined as $\Delta\rho=\rho
	_{E}-\rho_{0}$ where $\rho_{E}$ and $\rho_{0}$ are the charge densities of the
	bilayer MoSi$_{2}$N$_{4}$ and WSi$_{2}$N$_{4}$ with and without the external
	electric field, respectively. Here, positive (negative) $\Delta\rho$ indicates
	charge accumulation (depletion). As can be seen from the figure, there is
	charge accumulation in the bottom layer and depletion in the top layer. This
	leads to the VBM in the bottom layer shifting up and the CBM in the top layer
	shifting down, which reduces the energy band gap.
	
	In conclusion, we have shown that the biaxial strain and external electric
	field are effective ways for the band gap engineering of bilayer MoSi$_{2}%
	$N$_{4}$ and WSi$_{2}$N$_{4}$. Band gaps can be tuned by both compress and tensile
	strain. Compress strain can even leads to indirect band gap to direct band gap
	transition. The change in Mo/W-N bond length and Mo-Mo/W-W distance induced by
	strain is the main reason for this band structure engineering. Electric field
	can also modulate the band gap and even semiconductor to metal
	transition. Charge redistribution induced by the electric field accounts for
	the band gap tuning. Our study suggests the strain and electric field tunable
	bilayer MoSi$_{2}$N$_{4}$ and WSi$_{2}$N$_{4}$ could be promising materials for
	the next generation nanoelectronics and optoelectronics.
	
	This work is supported by Singapore MOE Tier 2 (Grant No. 2018-T2-1-007). All the
	calculations were carried out using the computational resources provided by
	the National Supercomputing Centre (NSCC) Singapore.
	
	\textbf{Data Availability} 
	
	The data that support the findings of this study are available from the corresponding author upon reasonable request.

\end{document}